\documentclass[twocolumn, times, showpacs,preprintnumbers,amsmath,amssymb,prl]{revtex4}

\usepackage{graphicx}

\usepackage{dcolumn}

\usepackage{bm}

\usepackage{times}

\begin{document}

\title{Magneto-Optical Trap for Thulium Atoms}

\author{D.~Sukachev}\altaffiliation[also at Moscow Institute of Physics and Technology,
141704 Dolgoprudny, Moscow reg. Russia]\
\author{A.~Sokolov}\altaffiliation[also at Moscow Institute of Physics and Technology,
141704 Dolgoprudny, Moscow reg. Russia]\
\author{K.~Chebakov}
\author{A.~Akimov}\altaffiliation[also at Moscow Institute of Physics and Technology,
141704 Dolgoprudny, Moscow reg. Russia]\
\author{S.~Kanorsky}
\author{N.~Kolachevsky}
\altaffiliation[also at Moscow Institute of Physics and Technology,
141704 Dolgoprudny, Moscow reg. Russia]\
\author{V.~Sorokin}\altaffiliation[also at Moscow Institute of Physics and Technology,
141704 Dolgoprudny, Moscow reg. Russia]\

 \affiliation{P.N. Lebedev Physical Institute, Leninsky prospekt 53, 119991 Moscow, Russia}




$^1$E-mail: kolik@lebedev.ru

\begin{abstract}
Thulium atoms are trapped in a magneto-optical trap using a strong
transition at 410 nm with a small branching ratio.  We trap up to
$7\times10^{4}$ atoms at a temperature of 0.8(2)\,mK after
deceleration in a 40 cm long Zeeman slower. Optical leaks from the
cooling cycle influence the lifetime of atoms in the  MOT which
varies between 0.3\,-1.5\,s in our experiments. The lower limit
for the leaking rate from the upper cooling level is measured to be
22(6)\,s$^{-1}$. The repumping laser transferring the atomic population out of the F=3
hyperfine ground-state sublevel gives a 30\% increase for the lifetime and
the number of atoms in the trap.
\end{abstract}

\maketitle

Laser cooling and trapping of neutral atoms is one of the most
powerful tools to study atomic ensembles at ultra-low temperatures~\cite{W.D.Phillips}.
It has opened a new era in  precision laser
spectroscopy~\cite{Riehle}, the study of collisions~\cite{Weiner},
atomic interferometry~\cite{Leggett} and the study of quantum
condensates~\cite{Ketterle_BEC, Regal}. Compared to buffer gas cooling~\cite{C.I.Hancox},
much lower temperatures may be reached, although laser cooling
strongly depends on an atomic level structure and the availability
of laser sources.

Since the very first experiments on laser deceleration of sodium
atoms~\cite{Balykin,Metcalf} alkalis  remain the most popular
objects for laser cooling. Today all alkali-earth atoms and
noble gases in the metastable state  (except Rn) are successfully
laser cooled and trapped in magneto-optical traps (MOTs).
The family of laser cooled species is continuously growing,
 with laser cooling and trapping of
 Er~\cite{McCelland}, Cd~\cite{Brickman}, Ra~\cite{Guest}, Hg~\cite{Hachisu} and Dy~\cite{Dy_cooling}
 being reported in the last few years. These
new cold species find applications in metrology, quantum
information, tests of fundamental theories,  and degenerative gas
studies.

We have demonstrated laser cooling and trapping of thulium
atoms in a MOT at a wavelength  of 410.6\,nm.
This lanthanide
possesses only one stable bosonic isotope, $^{169}$Tm, with
nuclear spin quantum number $I=1/2$.  After Yb~\cite{Honda},
Er~\cite{McCelland} and Dy~\cite{Dy_cooling} it is the fourth
lanthanide which has been trapped in
a MOT.  Lanthanides with unfilled $4f$ electronic shell
 are especially interesting due to the following peculiarities.

 First, their ground state magnetic moment is much larger than that of the alkalis.
The dipole moment of  $^{169}$Tm is $4\mu_B$ (here $\mu_B$ is the
Bohr magneton), for Er it approaches $7\mu_B$ and for Dy it is $10\mu_B$. Such strongly
magnetic atoms may be used in studies of dipole-dipole
interactions~\cite{Stuhler} and interactions with
superconductors~\cite{Cano}. Physics of
 cold polar molecules~\cite{Sawyer} may also benefit from using a strongly magnetic atom in molecular synthesis.

\begin{figure}[b!]
\centerline{
\includegraphics[width=0.45\textwidth]{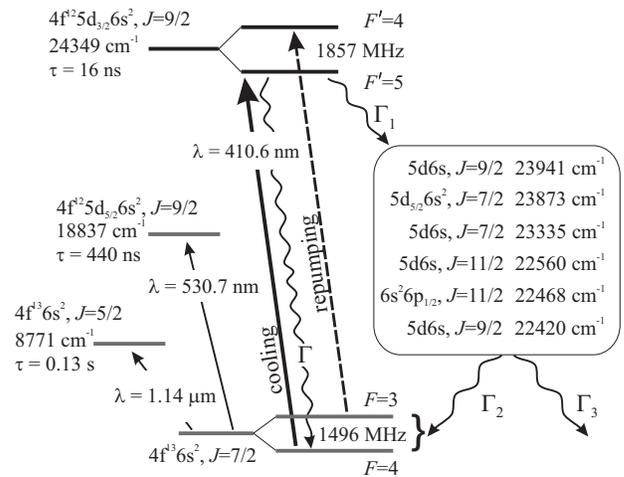}}
 \caption{Some relevant levels of atomic thulium. The transition $F=4\to F'=5$ is used for the laser cooling, while the
 transition $F=3\to F'=4$ is used for the repumping (optionally).
  The parameters $\Gamma$ and $\Gamma_{1,2,3}$ denote the corresponding decay rates.}\label{fig1}
\end{figure}

Second, the ground state $4f^{n}6s^2$ of such atoms is split into a
number of fine structure sublevels separated by large, up to
optical, frequency intervals. The fine structure of the thulium ground state
$4f^{13}6s^2$ consists of two sublevels with
the total electronic momentum quantum numbers
$J=7/2$ and $5/2$. This sublevels are optically coupled by
a narrow (width of 1.2(4)\,Hz~\cite{Kolachevsky})
magnetic dipole transition at 1.14\,$\mu$m (fig.\,\ref{fig1}). The upper sublevel being
a long-live metastable optically addressable state is potentially suitable
for applications in quantum memory~\cite{M.D.Eisaman, K.S.Choi, B.Zhao}.

Outer closed $6s^2$ and $5s^2$ shells strongly shield the transition at 1.14\,$\mu$m.
Experiments with magnetically trapped atoms~\cite{C.I.Hancox}
and with atoms implanted in a solid matrix~\cite{Ishikawa} have shown
a dramatic reducing of a sensitivity to collisions with noble  gases
and perturbations by electric fields. This transition
also may be used in the detailed study of Tm-Tm long range
quadrupole-quadrupole interactions~\cite{Doyle} and in the search
for the fine structure constant variation
($\dot\alpha/\alpha$) because of its quadratic $\alpha$-dependence~\cite{UFN}.

Formerly we have studied candidates for laser cooling  transitions
in thulium~\cite{Kolachevsky}. Theoretical analysis of the leak
rates from the cooling cycle indicated that the most favorable for
the first cooling step is $4f^{13}6s^{2}\, (J={7}/{2},\,F=4) \to
4f^{12}5d6s^{2}\, (J'={9}/{2},\,F'=5)$ transition at 410.6\,nm
with  the natural line width of
$\gamma=\Gamma/2\pi=10(4)$\,MHz, where small and capital letters denote rates in
Hz and cycles per second respectively. Here $F$ and $F'$ denote the
total atomic momentum quantum numbers for ground and excited
states respectively. In 2009 we demonstrated the Zeeman
deceleration of a Tm thermal beam using laser radiation at
410.6\,nm without a repumping laser~\cite{K.Chebakov}. The presence
of the decay channel from the upper cooling level
(fig.\,\ref{fig1}) did not prevent efficient deceleration, indicating
the feasibility of the further cooling and trapping of
atoms.

\begin{figure}[t!]
\centerline{
\includegraphics[width=0.49\textwidth]{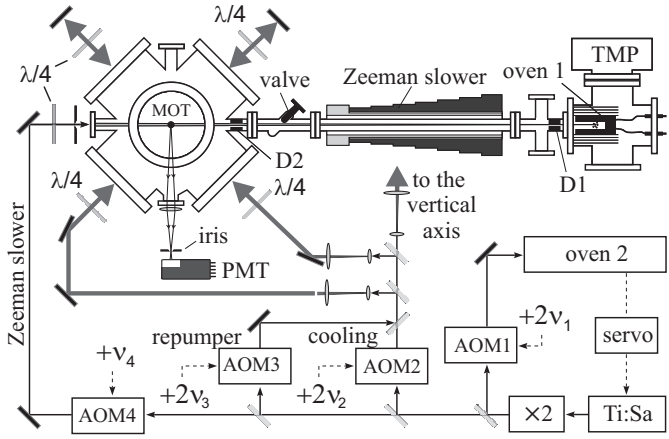}}
 \caption{The experimental setup. The PMT denotes a photomultiplier tube, the TMP -- a turbo-molecular pump. The 6-cross is pumped by a ion-getter pump (not shown).
 The AOMs 1,\,2,\,3 are working in a double-pass configuration at the frequencies $\nu_1=200$\,MHz, $\nu_2=190\div200$\,MHz, $\nu_3=300\div400$\,MHz, while the
 AOM4 is working at single pass at  $\nu_4=250$\,MHz.}\label{fig2}
\end{figure}

To trap Tm atoms we use a classical MOT configuration with three
orthogonal pairs of anti-propagating cooling laser beams as shown in
fig.\,\ref{fig2}. The MOT chamber is a 6-cross with additional ports
for a Zeeman slower and detectors. The chamber is pumped by a
30\,l/s ion-getter pump to a pressure less than $10^{-8}$\,mbar.
Two coils in the anti-Helmholtz configuration produce an axial
field gradient up to 10\,G/cm in the center of the chamber. The laboratory
magnetic field is compensated by additional coils.

The MOT is  loaded  from an atomic beam  decelerated in the Zeeman
slower. Tm vapors are produced in a home-made sapphire oven at a
temperature  1100\,K which is much lower than the  melting point
(1818\,K). This temperature provides a sufficient atomic flow from
the oven (the corresponding saturated vapor pressure is about
$10^{-2}$\,mbar). The oven is separately pumped by a 30\,l/s
turbo-molecular pump to $10^{-7}$\,mbar.

The atomic beam is formed by two cylindrical diaphragms: D1 (3\,mm
in diameter, 2\,cm long)  and D2 (5 \,mm in diameter, 1 cm long).
In the previous experiments~\cite{K.Chebakov}  the flow of slowed
atoms at the MOT position was measured to be
 $10^7$\,s$^{-1}$cm$^{-2}$ at a total flow (rate) of $10^9$\,s$^{-1}$cm$^{-2}$.
 The beam cross-section is 1\,cm$^2$ in the trapping region~\cite{K.Chebakov}.

Atoms are  cooled by the second harmonic of a Ti:sapphire laser at
410.6\,nm.  The laser frequency is stabilized to the Tm saturation
absorption signal from the separate oven 2. The second oven is  a
continuously pumped stainless steel tube with Tm chunks inside
which is heated to 900\,K~\cite{Kolachevsky,K.Chebakov}. The
acousto-optical modulator (AOM1) in a double-pass configuration
shifts the laser frequency by a fixed red detuning of 400\,MHz with
respect to the cooling transition.

Up to 20\,mW of the 410.6\,nm power is sent to the second AOM (AOM2), also
working in the double-pass configuration at around 200\,MHz.  This
produces the cooling beam with the desired red detuning which is
  then split in three nearly round
Gaussian shape beams of equal intensity, each expanded by a
separate telescope to a diameter of 5\,mm (at $1/e$ level). A
small fraction of power is sent to the third AOM (AOM3) which shifts the
frequency into resonance with the $F=3\to F'=4$ transition
 for repumping the atomic population from the
$F=3$ sublevel.

The fourth  single-pass AOM (AOM4)  works in the $+1$ order  to produce
the red-detuned frequency for the Zeeman slower. The frequency
detuning, the light power and the currents flowing through
Zeeman slower are optimized using the MOT luminescence signal. The
maximal number of trapped atoms is observed for a red detuning
of $150$\,MHz and the highest available light power (typically
about 15\,mW). At such detuning the Zeeman slower beam virtually
does not interact with trapped atoms. The slowing light tuned to
the resonance with the cooling transition also transfers the
population from the F = 3 sublevel to the cooling cycle due to
the off resonance excitation (the corresponding rate is about $10^{5}$ s$^{-1}$) thus
playing a role of the repumping laser in the atomic beam~\cite{McCelland,K.Chebakov}.

A fully loaded MOT has a typical size of 0.13\,mm in diameter (at
1/$e$ level) which is measured by a charge-coupled device (CCD).
In all other experiments the CCD was replaced by an absolutely
calibrated photo-multiplying tube (PMT) working in the current
measurement regime. The MOT luminescence at 410.6\,nm is collected
by a lens and is focused onto the plane of the photocathode. A iris
diaphragm of 3\,mm in diameter is placed in the image plane for
temperature measurements. The MOT image is adjusted to the iris
center.

The temperature of the atoms in the MOT is measured by the release and
recapture method described e.g. in~\cite{Lett, Cataliotti}.
The slower and the MOT beams are simultaneously
 switched off releasing atoms from the MOT. The MOT beams are switched on
 again after a certain time interval which was varied in steps from 1 ms to 30 ms.
 We measure the fraction of
atoms left in the registration area. 
This fraction is defined by the size of the iris diaphragm and the
atomic thermal velocity. For a red detuning of $\gamma$ and an intensity at the center of each beam $I_0=2$\,mW/cm$^2$  the measured temperature equals 0.8(2)K. This temperature is consistent with the evaluation obtained from the Doppler theory of optical molasses~\cite{Lett} which gives 1 mK for our experimental parameters. Equipartition theorem~\cite{Cd MOT} gives a similar result for the measured MOT size.  The Doppler cooling limit for the transition at 410.6 nm is $T_D=0.23$\,mK.
Lower temperatures may be achieved
by switching to another weaker cooling transition at 530.7\,nm~\cite{Kolachevsky}\,(see fig.\,\ref{fig1}).

 After the MOT beams are switched on again a quick ($\sim10$~ms)
 recapture  process takes place which is
 followed by the decay of the signal due to the trap losses.
  We observe a single exponential decay with no indication of magnetically
trapped atoms or refilling of the cooling cycle from the corresponding
population reservoir, as was measured for erbium~\cite{McCelland}. The MOT is reloaded in
3 seconds by switching on the slower light beam
(the steady-state MOT luminescence is not detectable in its absence).

\begin{figure}[t!]
\centerline{
\includegraphics[width=0.48\textwidth]{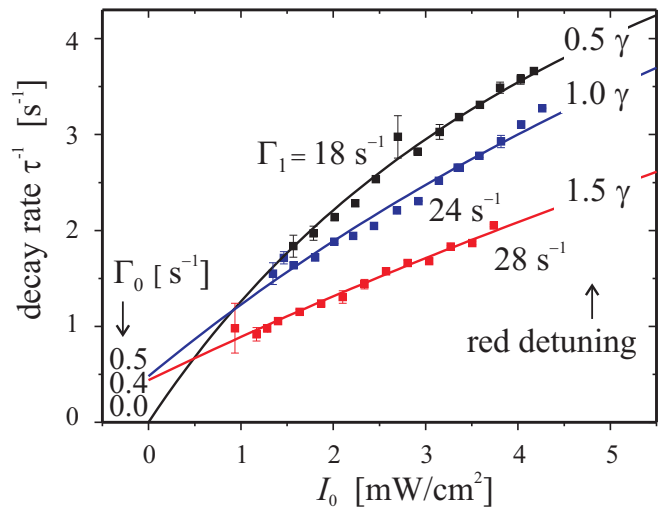}}
 \caption{ The decay time of trapped atom number measured for different intensities and detunings of
 the cooling beams. Squares show the experimental data obtained for
 the
 red detunings of $0.5\,\gamma$ (black), $\gamma$\,(blue) and $1.5\,
 \gamma$ (red). $I_0$ is the intensity on the axis of each of the 6
 MOT beams. Parameters $\Gamma_{0}$, $\Gamma_1$ correspond to the fits described in the
 text. The repumping laser is off.
    }\label{fig4}
\end{figure}

The lifetime of atoms in the MOT, $\tau$, is measured from the
decay curves by fitting an exponential to the
experimental data. At our atom number density of
$10^{11}$\,cm$^{-3}$ the role of binary collisions is negligible
and the MOT is optically thin which justifies the use of an exponential decay  fit.
We observe no systematic difference between the lifetimes deduced from the decay curves and from the loading curves.
At high intensities of the MOT beams ($I_{0} > 3$\,mW/cm$^2$, see fig.\,\ref{fig4}) collisions with atoms from the beam also play insignificant role. It was tested by  closing partially the valve (installed right after the slower, fig. 2) such way, that direct collisions with atoms from the beam are suppressed. In that case we observed three-fold reduction of the number of atoms, but the lifetime in the MOT remained unchanged.

Results for different on-axis intensities, $I_0$ (given per beam),
and detunings of the cooling beams are shown in fig.\,\ref{fig4}.
Reduction of the life time at higher intensity results from the
optical leaks to the six odd-parity levels shown in
fig.\,\ref{fig1}.

The complexity of configurations and a big number of intermediate
levels impede detailed theoretical analysis of the system. Here
we consider the simplest model of two cooling levels coupled with
the laser field with a decay channel from the upper level. We
assume that no population leaked from the cooling cycle returns
back via cascade decays to the ground state ($\Gamma_{2}=0$).
As experiments with the repumping laser indicate the
model is incomplete (see further), but it provides the lower limit for the
decay rate $\Gamma_{1}$. The extension of the model to a more
realistic situation (similar to that shown in fig.\,\ref{fig1}) gives
ambiguous results due to the lack of
experimental data.

The solution to the equations describing our model gives
\begin{equation}\label{eq1}
\tau^{-1}=\Gamma_{0}+\Gamma_{1} R/(1+R)\,,
\end{equation}
where parameter $\Gamma_{0}$ stands for losses independent of the
light intensity, e.g. collisions with a background gas. The
parameter $R$ is given by $R={S}/(1+S+4\kappa^2)$ where $\kappa$
is the detuning in units of $\gamma$ and $S=I_0/I_\textrm{sat}$ is
the saturation parameter ($I_\textrm{sat}=6.1$\,mW/cm$^2$).

The curves obtained by fitting (1) to the data as well as
the best fit parameters $\Gamma_{0}$ and $\Gamma_1$ are
shown in fig.\,\ref{fig4}. The decay rate $\Gamma_1$
grows with the increasing red detuning which is probably due to
the incompleteness of the model. The parameter $\Gamma_{0}$ obtained
for the detunings of $\gamma$ and $1.5\,\gamma$  corresponds to a
MOT lifetime of about 1.5\,s which is typical for other MOTs observed at similar vacuum conditions. For
the detuning of $0.5\,\gamma$ the transition is more strongly saturated  and
the extrapolation to zero intensity may result in a bigger
 error  for a given model. Moreover, at this detuning the
 MOT becomes very sensitive to the laser lock quality.   We conclude
that the lower limit for the decay rate from the level
$4f^{12}5d6s^{2}\, (J'={9}/{2})$ equals $22(6)$\,s$^{-1}$. It is
consistent with the previous theoretical estimation predicting a
rate between 300\,s$^{-1}$ and $1200$\,s$^{-1}$~\cite{Kolachevsky}.

\begin{figure}[t!]
\centerline{
\includegraphics[width=0.45\textwidth]{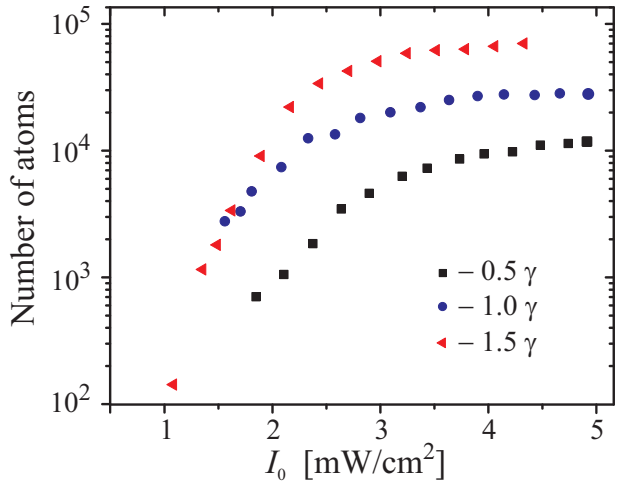}}
 \caption{The number of atoms in the MOT for different powers and red detunings of the cooling light.}
 \label{fig5}
\end{figure}

The number of atoms trapped in the MOT for different detunings and
powers of the cooling beams is shown in fig.\,\ref{fig5}.
The maximal number of atoms observed in our experiments corresponds
to $7\times 10^4$ for higher Zeeman slower and cooling
beam power and the number rapidly decreases for
 lower powers. The oven 1 temperature strongly influences the
number of atoms which indicates that the number of atoms in the
MOT is far from the saturation. An increase of the oven
temperature by 50\,K results in an approximately two-fold increase of
the signal. We deliberately did not increase the temperature to
avoid coating the Zeeman slower window.
The number of atoms may be further increased by implementation of the
``dark'' MOT\,\cite{Ketterle}.

\begin{figure}[t!]
\centerline{
\includegraphics[width=0.45\textwidth]{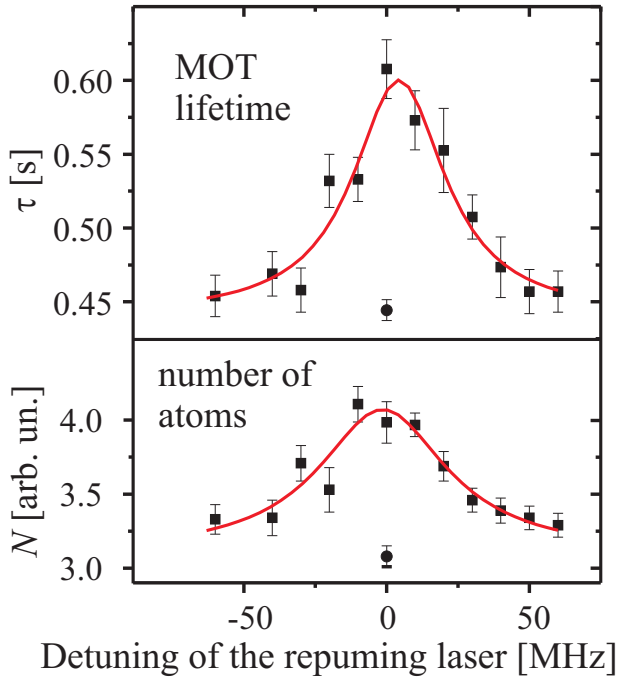}}
 \caption{The MOT life time $\tau$ and the number of atoms $N$ $\it vs.$ the repumping laser detuning
  from the $F=3\to F'=4$ resonance. The repumping beam intensity equals  2\,$\mu$W/cm$^2$ at the MOT position. The MOT beams red detuning is $\gamma$, $I_0=3$\,mW/cm$^2$. Solid
   curves are the Lorentzian fits. Data points denoted by circles are obtained without the repumping laser.}
\label{fig6}
\end{figure}

A presence of repumping laser tuned into exact resonance between
$F=3$ and $F'=4$ hyperfine sublevels (fig.\,\ref{fig1}) increases the life
time $\tau$ and the number of atoms $N$ in the MOT by about 30\% for a
given set of experimental conditions (see fig.\,\ref{fig6}).
This result indicates that a part
of population leaked from the cooling cycle returns back to the
ground state via further decays. Such a refilling channel
($\Gamma_2\neq0$) is not taken into account in the model
(\ref{eq1}) which can explain some peculiarities of the fit in
fig.\,\ref{fig4}. Unfortunately, the insufficient number of
observable parameters does not allow for qualitative characterization of
$\Gamma_2$. The repumping laser intensity of about 1\,$\mu$W/cm$^2$ is
enough to close the corresponding leak channel: fig.\,\ref{fig6}
corresponds to the
 saturated regime since the spectral width  of the fits
 is ~4 times broader than the natural line width.

In conclusion, we have demonstrated  laser cooling and trapping of
up to $7\times 10^4$ thulium atoms at 0.8(2)\,K in a
magneto-optical trap working at 410.6\,nm. Measurement of the lifetime
in the MOT gives a lower limit for the decay from the upper
cooling level $4f^{12}5d6s^{2}\, (J'={9}/{2})$ of
$22(6)$\,s$^{-1}$. The repumper laser is not obligatory, but
increases the lifetime and number of atoms in the MOT .

The work is supported by the RFBR grant 09-02-00649, Presidential
grant MD-3825.2009.2 and RSSF.

\end{document}